\documentclass[10pt,conference,letterpaper]{IEEEtran}
\usepackage{times,amsmath,epsfig}
\usepackage{latexsym}
\usepackage{comment}
\usepackage{lipsum}
\usepackage{color}
\usepackage{amssymb}
\usepackage{multicol}

\newcommand{\ignore}[1]{}
\makeatletter
\def\verbatim@font{\rmfamily\small}
\makeatother

\newcommand{\mc}[1]{\mathcal{ #1}}
\newcommand{\nit}[1]{{\it #1}}
\newcommand{\boxtheorem}{\hfill $\Box$\vspace*{0.2cm}}
\newcommand{\defproof}[2]{{\noindent\bf Proof of #1:\
}#2 \boxtheorem}

\newcommand{\red}[1]{\textcolor{red}{#1}}

\newcommand{\comlb}[1]{{\vspace{2mm}\noindent \bf \red{COMM(LEO):}}~ #1 \hfill {\bf
    END.}\\}

\newtheorem{example}{Example}

\newcounter{rownum}

\title{Extending Contexts with Ontologies for  Multidimensional Data Quality Assessment}
\author{%
{Mostafa Milani\ignore{{\small $~^{\#1}$}}, \ Leopoldo Bertossi\ignore{{\small $~^{\#2}$}}, \ Sina Ariyan\ignore{{\small $~^{\#3}$}}}%
\vspace{1.6mm}\\
\fontsize{10}{10}\selectfont\itshape
\ignore{$^{\#}$\,} Carleton University, School of Computer Science, Ottawa, Canada\\
\fontsize{9}{9}\selectfont\ttfamily\upshape
\vspace{1mm}\{mmilani,bertossi,mariyan\}@scs.carleton.ca\ignore{\\
$^{1}$\,mmilani@scs.carleton.ca\\
$^{2}$\,bertossi@scs.carleton.ca\\
$^{3}$\,mariyan@scs.carleton.ca}
\vspace{1.2mm}\\
\fontsize{10}{10}\selectfont\rmfamily\itshape
}
\begin{document}
\maketitle

\begin{abstract} ~Data quality and data cleaning are context dependent activities. Starting from this observation, in previous work a context model
for the assessment of the quality of a database instance was proposed. In that framework, the context takes the form  of a possibly virtual database or  data integration system into which a database instance under quality assessment is mapped, for additional analysis and processing, enabling quality assessment. In this work we extend contexts with dimensions, and by doing so, we make possible a multidimensional assessment of data quality assessment. Multidimensional contexts are represented as ontologies written in Datalog$\pm$. We use this language for representing {\em dimensional constraints}, and {\em dimensional rules}, and also for doing {\em query answering} based on dimensional navigation, which becomes an important auxiliary activity in the assessment of data. We show ideas and mechanisms by means of  examples.
\end{abstract}

% NOTE keywords are not used for conference papers so do not populate them
% \begin{keywords}
% keyword-1, keyword-2, keyword-3
% \end{keywords}
%
\vspace{-1mm}\section{Introduction}
\vspace{-1mm}The quality of data cannot be assessed without contextual knowledge about the production or the use of data. Actually, the notion of data quality is based on the degree in which the data fits or fulfills a form of usage \cite{BT06,JIA08}. As a consequence, the quality of data depends on their use context. It becomes clear that  context-based data quality assessment requires a formal model of context, at least for the use of data.

In this work we follow and extend the approach proposed in \cite{BR10}. According to it, the assessment of a database $\mc{D}$ is performed by mapping it into a context $\mc{C}$ that is represented as another database, or as a database schema with partial information, or, more generally, as a virtual data integration system with possibly some materialized data and access to external sources of data. The quality of data in $\mc{D}$ is determined through additional processing of data within the context. This process leads to a new (or possible several) quality version(s) of $\mc{D}$, whose quality is measured in terms of how much it departs from its quality version(s).

In \cite{BR10}, dimensions are not considered as contextual elements for data quality analysis. However, in practice dimensions are naturally associated to contexts. For example, in  \cite{TC13}, they become the basis for building contexts, and  in \cite{TRL10} they are used for data access data at query answering time.

In order to capture general dimensional aspects of data for inclusion in contexts, we take advantage of  the Hurtado-Mendelzon (HM) multidimensional data model \cite{HG05}, whose
inception was mainly motivated by data warehouse and OLAP applications. We extend and formalize it in ontological terms.
Actually, in \cite{MLK12} an extension of the HM model was proposed, with applications to data quality assessment in mind. That work was limited to a representation of this extension
in description logic (actually, an extension of DL-Lite \cite{DCAL07}), but data quality assessment was not developed.

In this work we propose an ontological representation in  Datalog$\pm$ \cite{ACL09-2} of the extended HM model, and also mechanisms for data quality assessment based on query answering from the ontology via dimensional navigation. Our extension of the HM model includes {\em categorical relations} associated to categories at different levels in the dimensional hierarchies, possibly to more than one dimension.
The extension also considers  {\em dimensional constraints} and {\em dimensional rules}, which could be treated both as {\em dimensional integrity constraints} on categorical relations that involve values from dimension categories.

However, dimensional constraints are intended to be used as {\em denial constraints} that forbid certain combinations of values, whereas the dimensional rules are intended to be used for data completion, to generate data through their enforcement. Dimensional constraints can be {\em intra-dimensional}, i.e. putting restrictions on data values of categorical relations associated to categories in a single dimension, or {\em inter-dimensional}, i.e. putting restrictions on data values of categorical relations associated to categories in different dimensions.

The next example illustrates the intuition behind categorical relations, dimensional constraints and rules, and how the latter can be used for data quality assessment. In it we assume, according to the HM model, that a dimension consists of a number of categories related to each other by a partial order. Later on, we use the example to show how contextual data can be captured as a Datalog$\pm$ ontology.

\begin{example}\label{exp:intr} Consider a relational table {\it Measurements} with body temperatures of patients in an institution (Table~\ref{tab:measurements}). A doctor in this institution needs the answer to the query: {\em ``The body temperatures of {\it Tom Waits} for {\it September 5} taken around noon with a thermometer of brand {\it B1}"}  (as he expected). It is possible that a nurse, unaware of this requirement, used a thermometer of brand {\it B2}, storing the measurements in {\it Measurements}. In this case, not all the measurements in the table are up to the expected quality. However, table  {\it Measurements} alone does not discriminate between expected or intended values (those taken with brand {\it B2}) and the others.

Now, for assessing the quality of the data in {\it Measurements} according to the doctor's quality requirement, extra contextual

 \newpage \vspace*{-5mm}\noindent information about the thermometers used may be useful. For instance, there is a table {\it PatientWard}, linked to the {\it Ward} category, that stores patients of each ward of the institution (Fig. \ref{fig:dim}). In addition, the institution has a {\em guideline}  prescribing that: {\em ``Temperature measurement for patients in standard care unit have to be taken with thermometers of Brand B1"}.

This guideline, which will become a dimensional rule in the ontology, can be used for data quality assessment when combined with an intermediate virtual relation, {\it PatientUnit},  linked to the {\it Unit} category, that is generated
 from  {\it PatientWard} by upward-navigation  through dimension \nit{Hospital} (on left-hand-side of Fig.~\ref{fig:dim}), from category \nit{Ward} to category \nit{Unit}.

 Now it is possible to conclude that on certain days, Tom Waits was in the standard care unit, where his temperature was taken, and with the right thermometer according to the guideline (patients in wards ${\it W_1}$ or ${\it W_2}$ had their temperatures taken with a thermometer of brand {\it B1}).
 These clean data appear  in relation $\nit{Measurements}^q$ (Table~\ref{tab:qualitymeasurements}), which can be seen as a quality answer to the doctor's request.

\ignore{
 \comlb{{\bf AFTER SKYPING WITH MOSTAFA, AND ASSUMING THAT THIS PAPER WILL TURN AROUND THIS FULLY DEVELOPED EXAMPLE.} Say that $\nit{Measurements}^q$ has to be defined on the basis
 of the original \nit{Measurements} and additional contextual information. The latter has to include information about thermometers (instruments) and their brands. This is what we
 did in the BIRTE paper. Assume for example (thinking aloud) that you have a table \nit{Taken(nurse,patient,thermometer,brand)}, then $\nit{Measurements}^q$ would be defined
 by something like this: $\nit{Measurements}^q(t,Waits,v) \leftarrow \nit{Measurements}(t,Waits,v), Taken(n,Waits,thermometer,B1)$.  Obtaining data for \nit{Taken} should be enabled
 by navigation as supported by Datalog, not hard-coded by the user as the second rule in example 6.}

 \comlb{If everything turns around the example, we should have the query posed to the ontology right in this example, and reformulated in terms of the quality predicate.}

 \comlb{It would be good to extend this example to show downward navigation, and a query that requires that. In my view, {\bf the essence of navigation (direction) is what we do in
 top-down approaches, when we go for the data we need, `a la Prolog} (even if we do bottom-up evaluation). This is important to keep in mind. }
}

\begin{table}
\begin{minipage}[t]{0.45\linewidth}\centering
\begin{center}
\setlength{\tabcolsep}{0.3em}
\setcounter{rownum}{0}
\setlength{\arrayrulewidth}{0.75pt}
\renewcommand*\arraystretch{1.20}
\caption{{\it Measurements}}
\label{tab:measurements}
\begin{tabular}{c|c|c|c|}
\cline{2-4}
 & \textbf{Time} & \textbf{Patient} & \textbf{Value}\\
\cline{2-4}
{\tiny \addtocounter{rownum}{1}\arabic{rownum}} & Sep/5-12:10 & Tom Waits & 38.2 \\
\cline{2-4}
{\tiny \addtocounter{rownum}{1}\arabic{rownum}} & Sep/6-11:50 & Tom Waits & 37.1 \\
\cline{2-4}
{\tiny \addtocounter{rownum}{1}\arabic{rownum}} & Sep/7-12:15 & Tom Waits & 37.7 \\
\cline{2-4}
{\tiny \addtocounter{rownum}{1}\arabic{rownum}} & Sep/9-12:00 & Tom Waits & 37.0 \\
\cline{2-4}
{\tiny \addtocounter{rownum}{1}\arabic{rownum}} & Sep/6-11:05 & Lou Reed & 37.5 \\
\cline{2-4}
{\tiny \addtocounter{rownum}{1}\arabic{rownum}} & Sep/5-12:05 & Lou Reed & 38.0 \\
\cline{2-4}
\end{tabular}
\end{center}
\end{minipage}
\begin{minipage}[t]{0.45\linewidth}\centering
\begin{center}
\setlength{\tabcolsep}{0.3em}
\setcounter{rownum}{0}
\setlength{\arrayrulewidth}{0.75pt}
\renewcommand*\arraystretch{1.20}
\caption{${\it Measurements}^q$}
\label{tab:qualitymeasurements}
\begin{tabular}{c|c|c|c|}
\cline{2-4}
 & \textbf{Time} & \textbf{Patient} & \textbf{Value}\\
\cline{2-4}
{\tiny \addtocounter{rownum}{1}\arabic{rownum}} & Sep/5-12:10 & Tom Waits & 38.2 \\
\cline{2-4}
{\tiny \addtocounter{rownum}{1}\arabic{rownum}} & Sep/6-11:50 & Tom Waits & 37.1 \\
\cline{2-4}
\end{tabular}
\end{center}
\end{minipage}
\end{table}

\begin{figure}
\begin{center}\vspace{-3mm}
 \psfig{figure=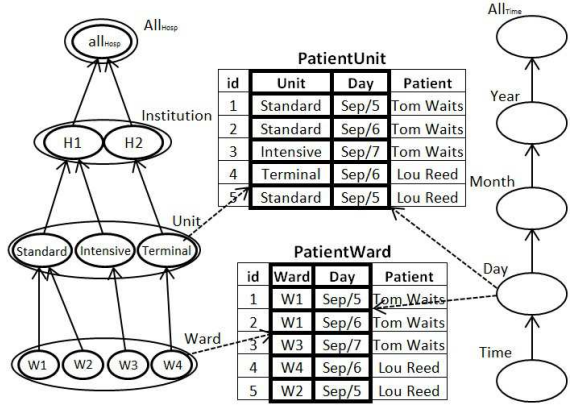,width=75mm}
 \caption{An extended  multidimensional model}\label{fig:dim}
\end{center} \vspace{-6mm}
\end{figure}

Elaborating on this example, it could be the case that there is a {\em constraint} imposed  on dimensions and relations linked to their categories. For instance, one capturing that the intensive care unit was closed since August/2005: {\em ``No patient was in intensive care unit during the time after August /2005"}. Again, through upward-navigation to the next category, we should conclude that the third tuple in table  {\it PatientWard} should be discarded. This {\em inter-dimensional constraint} involves dimensions {\it Hospital} and {\it Time}  (right-hand-side of Fig. \ref{fig:dim}), to which the ward and the day values in {\it PatientWard} are linked.
\boxtheorem\end{example}

The example shows a processing of data that involves changing the level of data linked to a dimension. This form of  {\em dimensional navigation} may be required for query answering both in the {\em downward} and {\em upward} directions  (Example~\ref{exp:intr} shows the latter). Our ontological multidimensional contexts support both. \ignore{The next example shows the need for downward navigation when generating intensional data in categorical relations for query answering.}

\begin{example}[ex. \ref{exp:intr} cont.]\label{exp:downward}
 Two additional categorical relations, {\it WorkingSchedules} and {\it Shifts} (Table~\ref{tab:ws} and Table~\ref{tab:shifts}), store shifts of nurses in wards and schedules of nurses in units. A query to {\it Shifts} asks for dates when {\it Mark} was working in ward {\it W2}, which has no answer with the extensional data in Table~\ref{tab:shifts}. Now, an institutional guideline states that if a nurse works in a unit on a specific day, he/she has shifts in every ward of that unit on the same day. Consequently, the last tuple in Table~\ref{tab:ws} implies that {\it Mark} has shifts in both {\it W1} and {\it W2} on {\it Sep/9}. This date would be an answer obtained via downward navigation from the {\it Standard} unit to its wards (including {\it W2}). \boxtheorem
\end{example}

\vspace*{-0.7cm}
%\begin{center}
\begin{table}[h]
\hspace*{8mm}\begin{minipage}[t]{0.45\linewidth}\centering
\setcounter{rownum}{0}
\setlength{\tabcolsep}{0.3em}
\setlength{\arrayrulewidth}{0.75pt}
\renewcommand*\arraystretch{1.15}
\caption{\it WorkingSchedules}
\label{tab:ws}
\vspace{-3mm}\begin{tabular}{c|c|c|c|c|}
\cline{2-5}
 & \textbf{Unit} & \textbf{Day} & \textbf{Nurse} & \textbf{Type}\\
\cline{2-5}
{\tiny \addtocounter{rownum}{1}\arabic{rownum}} & Intensive & Sep/5 & Cathy & cert. \\
\cline{2-5}
{\tiny \addtocounter{rownum}{1}\arabic{rownum}} & Standard & Sep/5 & Helen & cert. \\
\cline{2-5}
{\tiny \addtocounter{rownum}{1}\arabic{rownum}} & Standard & Sep/6 & Helen & cert. \\
\cline{2-5}
{\tiny \addtocounter{rownum}{1}\arabic{rownum}} & Terminal & Sep/5 & Susan & non-c. \\
\cline{2-5}
{\tiny \addtocounter{rownum}{1}\arabic{rownum}} & Standard & Sep/9 & Mark & non-c. \\
\cline{2-5}
\end{tabular}
\end{minipage}
\begin{minipage}[t]{0.45\linewidth}\centering
\setlength{\tabcolsep}{0.3em}
\setcounter{rownum}{0}
\setlength{\arrayrulewidth}{0.75pt}
\renewcommand*\arraystretch{1.20}
\caption{{\it Shifts}}
\label{tab:shifts}
\vspace{-0.2cm}

\begin{tabular}{c|c|c|c|c|}
\cline{2-5}
 & \textbf{Ward} & \textbf{Day} & \textbf{Nurse} & \textbf{Shift}\\
\cline{2-5}
{\tiny \addtocounter{rownum}{1}\arabic{rownum}} & W4 & Sep/5 & Cathy & night \\
\cline{2-5}
{\tiny \addtocounter{rownum}{1}\arabic{rownum}} & W1 & Sep/6 & Helen & morning \\
\cline{2-5}
{\tiny \addtocounter{rownum}{1}\arabic{rownum}} & W4 & Sep/5 & Susan & evening \\
\cline{2-5}
\end{tabular}
\end{minipage}
\end{table}
%\end{center}
\vspace*{-0.4cm}

Example~\ref{exp:downward} shows that downward navigation is necessary for query answering, in this case, for propagating data in {\it WorkingSchedules} at the {\it Unit} level down to {\it Shifts} at the lower {\it Ward} level). In this process a unit may drill-down to more than one ward, e.g. {\it Standard} unit is connected to wards {\it W1} and {\it W2}), generating more than one tuple in {\it Shifts}.

Contexts should be represented as formal theories into which other objects, such as database instances, are mapped into, for contextual analysis, assessment, interpretation, additional processing, etc. \cite{BR10}. Consequently,  we show how to represent multidimensional contexts as logic-based ontologies (c.f. Section \ref{sec:dlrep}). These ontologies represent and extend the HM multidimensional model (cf. Section~\ref{sec:preliminaries}). Our ontological language of choice is Datalog$\pm$ \cite{ACL12}. It allows us to give a clear semantics to our ontologies, to support some forms of logical reasoning, and to apply some query answering algorithms. Furthermore, Datalog$\pm$ allows us to generate explicit data by completion where they are missing, which is particularly useful for data generation though dimensional navigation.

Our ultimate goal is to use multidimensional ontological contexts for data quality assessment \cite{BR10}, which is achieved by introducing and defining in the context relational predicates standing for the {\em quality versions of relations in the original  instance}. Their definitions use additional conditions on data, to make them contain quality data. In this work, going beyond \cite{BR10}, the context also contains an ontology in Datalog$\pm$ that represents all the multidimensional elements shown in the examples above.

Our ontologies fall in the  {\em weakly-sticky} (WS) class \cite{ACL12} of the Datalog$\pm$ family of languages \cite{ACL09-2} (cf. Section \ref{sec:dlrep}) with {\em separable} equality generating dependencies (when used as dimensional constraints), which  guarantees that conjunctive query answering can be done in polynomial time in data. We have developed and implemented a deterministic algorithm for boolean conjunctive query answering, which is based on a non-deterministic algorithm  for WS Datalog$\pm$ \cite{ACL12}. The algorithm can be used with ontologies containing  dimensional rules that support both upward or downward navigation (cf. Section \ref{sec:qa}). Section \ref{sec:cdqa} shows how to  use the ontology to populate the quality versions of original relations.

This paper is an extended abstract. We show concepts, ideas, ontologies, and mechanisms only by means of an extended example. The general approach and its analysis
in detail will be presented in an extended version of this work.

\vspace{-2mm}
\section{Preliminaries}\label{sec:preliminaries}

\vspace{-1mm}
 We start from the HM multidimensional (MD) data model \cite{HG05}. In it, dimensions represent the hierarchical data; and facts describe  data as points in an MD space.
A \nit{dimension} is composed of a schema and an instance. A \nit{dimension schema} includes a directed acyclic graph (DAG) of {\em categories}, which defines {\em levels} of the category hierarchy. A dimension hierarchy corresponds to a partial-order relation between the categories, a so-called \nit{parent-child relation}. A \nit{dimension instance} consists of set of members for each category. The instance hierarchy corresponds to a partial-order relation between
members of categories, that parallels the {\em parent-child} relation between categories.  {\it Hospital} and {\it Time}, at the right- and left-hand sides of Fig. \ref{fig:dim}, resp., are dimensions.

We extend the HM model with, among other elements, {\em categorical relations}, which can be seen as a generalization of fact tables, but at different dimension levels and not necessarily containing numerical data.
Categorical relations represent the entities associated to the factual data. A \nit{categorical relation} has a schema and an instance. A \nit{categorical relation schema} is composed of a relation name and a list of attributes. Each attribute is either {\em categorical} or {\em non-categorical}. A categorical attribute takes as values the members of a category in a dimension. A non-categorical attribute takes values from an arbitrary domain.

\begin{example}[ex.~\ref{exp:intr} cont.]\label{exp:crelation}
In Fig.~\ref{fig:dim}, the categorical relation ${\it PatientWard(Ward, Day, Patient)}$ has its categorical attributes, {\it Ward} and {\it Day},  connected to the {\it Hospital} and {\it Time} dimensions.  {\it Patient} is a non-categorical attribute with patient names as values (there could be a foreign key to another categorical relation that stores data of patients).\boxtheorem
\end{example}
Datalog$\pm$~\cite{ACL09-2} is a family of languages that extends plain Datalog with additional elements: (a)  existential quantifiers in heads of
{\em tuple-generating dependencies} (TGDs); (b) {\em equality-generating dependencies} (EGDs), that use equality in heads; and (c) {\em negative constraints}, that use
$\bot$ in heads.
With these extensions, Datalog$\pm$  captures ontological knowledge that cannot be expressed in classical Datalog.

 Although the {\em chase} with these rules does not necessarily terminate, syntactic restrictions imposed on the set of rules aim to ensure decidability of conjunctive query answering, and is some cases, also tractability  in data complexity. Datalog$\pm$ has sub-languages, such as {\em linear}, {\em guarded}, {\em weakly-guarded}, {\em sticky}, and {\em weakly-sticky}, that depend on the kind of predicates and syntactic interaction of TGD rules that appear in the Datalog$\pm$ program.

In this paper, our MD ontologies turn out to be written in {\em weakly-sticky} (WS) Datalog$\pm$. This sublanguage extends {\em sticky} Datalog$\pm$ \cite{ACL10-2}. WS Datalog$\pm$ allows joins in the body of  TGDs, but with a milder restriction on the repeated variables. Boolean conjunctive query answering is tractable for WS Datalog$\pm$ \cite{ACL10-2}.

\vspace{-1mm}
\section{The Extended MD Model in Datalog$\pm$}\label{sec:dlrep}

\vspace{-1mm}
We will represent our extended MD model as a Datalog$\pm$ ontology  $\mc{M}$ that contains a schema $\mc{S}_\mc{M}$, an instance $\mc{D}_\mc{M}$, and a set of dimensional rules and constraints $\Sigma_\mc{M}$.
\ $\mc{S}_\mc{M} = \mc{K} \cup \mc{O} \cup \mc{R}$ is a finite set of predicates (relation names), where $\mc{K}$ is a set of {\em category predicates} (unary predicates), $\mc{O}$ is a set of {\em parent-child predicates}, i.e. partial-order relations between elements of adjacent categories, and $\mc{R}$ is a set of  {\em categorical predicates}.
In Example \ref{exp:intr}, $\nit{K}$ contains, e.g.  $\nit{Ward}(\cdot), \nit{Unit}(\cdot)$; $\mc{O}$ contains, e.g. a predicate for connections from \nit{Ward} to \nit{Unit};
 and $\mc{R}$ contains, e.g. \nit{PatientWard}.
An {\em instance}, $\mc{D}_\mc{M}$,  is a relational instance that gives (possibly infinite) extensions to the predicates in  $\mc{S}_\mc{M}$, and satisfies a given set of TGDs, EGDs, and negative constraints $\Sigma_\mc{M}$ (cf. below). The constants for $\mc{D}_\mc{M}$ come from an infinite underlying domain.

The dimensional rules and constraints in $\Sigma_\mc{M}$ constitute the intentional part of  $\mc{M}$. Rules (\ref{frm:gf1})-(\ref{frm:gf4}) below show the general form of elements of  $\Sigma_\mc{M}$. In what follows,  each $R_i(\bar{e}_i;\bar{a}_i)$ is a categorical atom, with $\bar{e}_i$ a sequence of categorical attributes (values) and $\bar{a}_i$ a sequence of non-categorical attributes; $D_i(e_i,e'_i)$ is a parent-child atom with $e_i,e'_i$  parent/child elements, resp.; and $K_i(e_i)$ is a category atom, with $e_i$ a category element. That is, $K_i \in \mc{K}, D_i \in \mc{O}$, $R_i \in \mc{R}$. As an instance in (\ref{frm:exref}) and (\ref{frm:exegd}), ${\it Unit}(u)$ is a category atom and ${\it UnitWard}(u,w)$ is a parent-child atom.
\begin{itemize}
\item [(a)] To capture the {\em referential constraint} between a categorical attribute of a categorical relation and a category, we use a negative constraint, with $e \in \bar{e}_i$:\footnote{Alternatively, we could have referential constraints between categorical relations and categories that are captured by Datalog$\pm$ TGDs, making it possible to generate elements in  categories or categorical relations.} \vspace{-1mm}
\begin{align}
\bot \  &\leftarrow~ R_{i}(\boldsymbol{\bar{e}_i};\bar{a}_i),\lnot K(e).\label{frm:gf1}
\end{align}\vspace{-6mm}
\item [(b)]
A {\em dimensional constraint} is  either an EGD of the form (\ref{frm:gf2}) (where $x,x'$ also appear in the body) or a negative constraint of the form (\ref{frm:gf3}): \vspace{-2mm}
\begin{align}
x=x' ~\leftarrow~&  R_i(\boldsymbol{\bar{e}_i};\bar{a}_i),...,R_j(\boldsymbol{\bar{e}_j};\bar{a}_j), \label{frm:gf2}\\
&D_n(e_n,e'_n),...,D_m(e_m,e'_m).\nonumber \vspace{-3mm}\\
\bot ~\leftarrow~& R_i(\boldsymbol{\bar{e}_i};\bar{a}_i),...,R_j(\boldsymbol{\bar{e}_j};\bar{a}_j), \label{frm:gf3}\\
&D_n(e_n,e'_n),...,D_m(e_m,e'_m). \nonumber  \vspace{-4mm}
\end{align}
\vspace{-6mm}
\item [(c)]
A {\em dimensional rule} is a Datalog$\pm$ TGD of the  form: \vspace{-2mm}
\begin{eqnarray}\label{frm:gf4}
\exists \bar{a}_z\;R_k(\bar{e}_k;\bar{a}_k)\ ~\leftarrow& \hspace{-4mm}R_i(\boldsymbol{\bar{e}_i};\bar{a}_i),...,R_j(\boldsymbol{\bar{e}_j};\bar{a}_j),\\
&  D_n(e_n,e'_n),...,D_m(e_m,e'_m).\nonumber
\end{eqnarray}
\vspace{-6mm}\phantom{poto}
\\Here, $\bar{a}_z \subseteq \bar{a}_k$, $\bar{e}_k \subseteq \bar{e}_i \cup ... \cup \bar{e}_j \cup \{e_n,...,e_m\} \cup \{e'_n,...,e'_m\}$ and $\bar{a}_k \! \smallsetminus \! \bar{a}_z \subseteq \bar{a}_i \cup ... \cup \bar{a}_j$. Furthermore, shared variables in bodies of TGDs correspond only to categorical attributes of categorical relations.
\end{itemize}
 With rule  (\ref{frm:gf4}) (an example is (\ref{frm:upward}) below), the possibility of doing dimensional navigation is captured by  joins between categorical predicates, e.g. $R_i(\boldsymbol{\bar{e}}_i;\bar{a}_i),...,R_j(\boldsymbol{\bar{e}}_j;\bar{a}_j)$ in the body, and parent-child predicates, e.g. $D_n(e_n,e'_n),...,D_m(e_m,e'_m)$.
Rule~(\ref{frm:gf4}) allows navigation in both upward and downward directions. The {\em direction of navigation} is determined by the level of categorical attributes that participate in the join  in the body. Assuming the join  is between $R_i(\bar{e}_i;\bar{a}_i)$ and $D_n(e_n,e'_n)$, upward navigation is enabled when $e'_n \in \bar{e}_i$ (i.e. $e'_n$ appears in $R_i(\bar{e}_i;\bar{a}_i)$) and $e_n \in \bar{e}_k$ (i.e $e_n$ appears in the head). On the other hand, if $e_n$ occurs in $R_i$ and $e'_n$ occurs in $R_k$, then downward navigation is enabled,  from $e_n$ to $e'_n$.

The existential variables in (\ref{frm:gf4}) make up for missing non-categorical attributes due to different schemas (i.e. the existential variables may appear in positions of non-categorical attributes but not in categorical attributes). As a result, when drilling  down, for each tuple of a categorical relation linked to a parent member, the rule generates tuples for all the child members of the parent member (or children specifically indicated in the body).

\begin{example}[ex.~\ref{exp:crelation} cont.] \label{exp:ont} The categorical attribute \nit{Unit} in categorical relation {\it PatientUnit} takes values from the {\it Unit} category. We use a constraint of the form (\ref{frm:gf1}). Similar constraints are in the ontology that capture the connection between other categorical relations and their corresponding categories. \vspace{-2mm}
\begin{align}\label{frm:exref}
\bot &~\leftarrow~ {\it PatientUnit(\boldsymbol{u},\boldsymbol{d};p)},\lnot {\it Unit}(u).
\end{align}
\vspace{-6mm}\phantom{poto}\\For the constraint in Example~\ref{exp:intr} requiring {\it ``No patient was in intensive care unit during the time after August 2005"}, we use a dimensional constraint of the  form (\ref{frm:gf3}): \vspace{-2mm}
\begin{align}
\bot ~\leftarrow~& {\it PatientWard(\boldsymbol{w},\boldsymbol{d};p)},{\it UnitWard}({\tt Intensive}, w),\nonumber\\
&{\it MonthDay}({\tt August/2005},d).\nonumber
\end{align}
\vspace{-7mm}\phantom{poto}\\Similarly, the following rule, of form (\ref{frm:gf2}), states that {\em ``All the thermometers used in a unit are of the same type"}: \vspace{-2mm}
\begin{align}
t=t' ~\leftarrow~& {\it Thermometer(\boldsymbol{w},\boldsymbol{t};n)},{\it Thermometer(\boldsymbol{w'},\boldsymbol{t'};n')},\nonumber\\
&{\it UnitWard(u,w)},{\it UnitWard(u,w')},\label{frm:exegd}
\end{align}
\vspace{-6mm}\phantom{poto}\\with ${\it Thermometer(Ward,Thermometertype;}$$\nit{Nurse})$ a categorical relation with thermometers used by nurses in wards.

Finally, the following dimensional rules of the form (\ref{frm:gf4}) capture how data in \nit{PatientWard} and \nit{WorkingSchedules} generate data for \nit{PatientUnit} and \nit{Shifts}, resp.:\footnote{A rule with a conjunction in the head can be transformed into a set of rules with single atoms in heads.} \vspace{-2mm}
\begin{align}
{\it PatientUnit(\boldsymbol{u},\boldsymbol{d};p)} ~\leftarrow~& {\it PatientWard(\boldsymbol{w},\boldsymbol{d};p)},\label{frm:upward}\\
&{\it UnitWard(u,w)}. \nonumber
\end{align}
\vspace*{-0.9cm}
\begin{align}
\exists z\;{\it Shifts(\boldsymbol{w},\boldsymbol{d};n,z)} ~\leftarrow~& {\it WorkingSchedules(\boldsymbol{u},\boldsymbol{d};n,t)},\nonumber\\
&{\it UnitWard(u,w)}. \label{frm:downward1}
\end{align}
\vspace{-6mm}\phantom{poto}

In (\ref{frm:upward}), dimension navigation is enabled by the join between {\it PatientWard} and {\it UnitWard}. The rule generates data for {\it PatientUnit} (at a the higher level of {\it Unit}) from {\it PatientWard} (at the lower level of {\it Ward}) via upward navigation. Notice that (\ref{frm:upward}) is in the general form (\ref{frm:gf4}), but since in this case the schemas of the two involved categorical relations match,  no existential quantifiers are necessary.

Rule~(\ref{frm:downward1}) captures downward navigation while it generates data for {\it Shifts} (at the level of {\it Ward}) from {\it WorkingSchedules} (at the level of {\it Unit}). In this case, the schemas of the two categorical relations do not match. So, the existential variable $z$ represents missing data for the {\it shift} attribute.\boxtheorem
\end{example}

It is possible to verify that {\em the Datalog$\pm$ MD ontologies with rules of the forms (\ref{frm:gf1})-(\ref{frm:gf4}) are weakly-sticky}. This follows from the fact that shared variables in the body of dimensional rules, as defined in (\ref{frm:gf4}), may occur only in positions of categorical attributes, where only limited values may appear, which depends on the assumption that the MD ontology has a fixed dimensional structure, in particular, with a fixed number of category members. No new category member is generated when applying the dimensional rules of the form (\ref{frm:gf4}).

The {\it separability property}~\cite{ACL10-1,ACL12} in relation to the interaction of dimensional EGDs of the form (\ref{frm:gf2}) and TGDs of the form (\ref{frm:gf4}) must be checked independently. However, {\em when the EGDs have only categorical variables in the heads, the separability condition holds}, which is the case with rule (\ref{frm:exegd}).

\begin{example}[ex.~\ref{exp:downward} and \ref{exp:ont} cont.]
To illustrate  query answering via downward navigation, reconsider the query about the dates that {\it Mark} works in {\it W1}:
$\mathcal{Q}'(d)\leftarrow {\it Shifts}({\tt W1}, d, {\tt Mark}, s)$.
Considering (\ref{frm:downward1}) and the last tuple in {\it WorkingSchedules}, the chase will generate a new tuple in {\it Shifts} for {\it Mark} on {\it Sep/9} in {\it W2}, with a fresh null value for his shift, reflecting incomplete knowledge about this attribute at the lower level. So, the answer to the query via (\ref{frm:downward1}) is {\it Sep/9}. \boxtheorem
\end{example}

\vspace{-2mm}The general TGD (\ref{frm:gf4}) only captures downward navigation when there is incomplete data about the values of non-categorical attributes, because existential variables are only non-categorical.
However, in some cases we may have incomplete data about the categorical attributes, i.e. about parents and children involved in downward navigation.

\vspace{-3mm}
\begin{table}[h]
\begin{center}
\setcounter{rownum}{0}
\setlength{\tabcolsep}{0.3em}
\setlength{\arrayrulewidth}{0.75pt}
\renewcommand*\arraystretch{1.15}
\caption{\it DischargePatients}\vspace{-2mm}
\label{tab:discharge}
\begin{tabular}{c|c|c|c|}
\cline{2-4}
 & \textbf{Inst.} & \textbf{Day} & \textbf{Patient}\\
\cline{2-4}
{\tiny \addtocounter{rownum}{1}\arabic{rownum}} & H1 & Sep/9 & Tom Waits \\
\cline{2-4}
{\tiny \addtocounter{rownum}{1}\arabic{rownum}} & H1 & Sep/6 & Lou Reed \\
\cline{2-4}
{\tiny \addtocounter{rownum}{1}\arabic{rownum}} & H2 & Oct/5 & Elvis Costello \\
\cline{2-4}
\end{tabular}
\end{center}
\end{table}

\vspace{-6mm}
\begin{example}[ex. \ref{exp:intr} cont.]
There is an additional categorical relation {\it DischargePatients} (Table~\ref{tab:discharge}) with data about patients leaving an institution.
Since each of them was in exactly one of the units,
{\it DischargePatient} should generate data for {\it PatientUnit} through downward navigation from the {\it Institution} level to the {\it Unit} level. Since we do not have knowledge about which unit at the lower level has to be specified, the following rule could be used: \vspace{-2mm}
\begin{align}
\exists u\;{\it InstitutionUnit(i,u)},&{\it PatientUnit(\boldsymbol{u},\boldsymbol{d};p)} ~\leftarrow~\label{frm:downward2}\\
 &{\it DischargePatients(\boldsymbol{i},\boldsymbol{d};p)}, \nonumber
\end{align}
\vspace{-7mm}\phantom{poto}\\which is not of the form (\ref{frm:gf4}), because it has an existentially quantified categorical variable, $u$, for units. It allows downward navigation while capturing incomplete data about units,
 and represents disjunctive knowledge at level of units. \boxtheorem
\end{example}

The general form of (\ref{frm:downward2}), for this type of downward navigation is as follows: \vspace{-2mm}
\begin{align}\label{frm:gf5}
\exists \bar{z}\;R_k(\bar{e}_k;\bar{a}_k),D_n(e_n,e'_n)&,...,D_m(e_m,e'_m) \ \ \leftarrow\\
&  R_i(\boldsymbol{\bar{e}_i};\bar{a}_i),...,R_j(\boldsymbol{\bar{e}_j};\bar{a}_j),\nonumber
\end{align}
\vspace{-6mm}\phantom{poto}\\
 where $\bar{z} \subseteq \bar{e}_k \cup \bar{a}_k \cup \{e_n,...,e_m\} \cup \{e'_n,...,e'_m\}$ and $\bar{e}_k \cup \{e_n,...,e_m\} \cup \{e'_n,...,e'_m\} \! \smallsetminus \! \bar{z} \subseteq \bar{e}_i \cup ... \cup \bar{e}_j$ and $\bar{a}_k \! \smallsetminus \! \bar{z} \subseteq \bar{a}_i \cup ... \cup \bar{a}_j$, and the categorical attributes of $R_i,\ldots,R_j$ refer to categories that are at a higher or same level than the categorical
 attributes
 of $R_k$. (In (\ref{frm:downward2}),  categories $\nit{Institution}$ and $\nit{Day}$ for $\nit{DischargePatients}$ are higher and same level, resp. than $\nit{Unit}$ and $\nit{Day}$ for $\nit{PatientUnit}$.)

 {\em If the MD ontology also includes rules of the form (\ref{frm:gf5}), it still is weakly-sticky}. This is because, despite the fact that these rules may generate new members (nulls), they can only generate a limited number of such members (because the rule only navigates in downward direction), i.e. there is no cyclic behavior. With these new rules, EGDs with only categorical attributes in heads do not guarantee separability anymore. So, checking
 this condition becomes application dependent.

\vspace{-2mm}
\section{Query Answering on MD Ontologies}\label{sec:qa}

\vspace{-1mm}
Weakly-stickyness guarantees that {\em boolean conjunctive query answering from our MD contextual ontologies becomes tractable} in data complexity \cite{ACL12}. Then, answering open conjunctive queries from the MD ontology is also tractable \cite{FG03}.

We have developed and implemented a deterministic algorithm, {\tt DeterministicWSQAns}, for answering boolean conjunctive queries from Datalog$\pm$ MD contextual ontologies. The algorithm is based on a non-deterministic algorithm, {\tt WeaklyStickyQAns}, for WS Datalog$\pm$ that runs in polynomial time in the size of extensional database \cite{ACL12}.

Given a set of WS TGDs, a boolean conjunctive query, and an extensional database, {\tt WeaklyStickyQAns} builds an ``accepting resolution proof schema", a tree-like structure which shows how query atoms can be entailed from the extensional instance. The algorithm rejects if there is no resolution proof schema; otherwise it builds it and accepts.

Our deterministic algorithm, {\tt DeterministicWSQAns}, applies a top-down backtracking search for accepting resolution proof schemas. Starting from the query, the algorithm resolves the atoms of the query, from left to right. In each step, an atom is resolved either by finding a substitution that maps the atom to a ground atom in the extensional database (which makes a leaf node) or by applying a TGD rule that entails the atom (building a subtree). The decision at each step is stored on a stack to be restored later if the algorithm fails to entail the atoms of the query in the next steps. The algorithm accepts if it resolves all the atoms in the query (the content of the stack specifies the decisions that lead to the accepting resolution proof schema), and rejects if it cannot resolve an atom, no matter what decisions have been made before.

In this deterministic approach, possible substitutions of constants for query variables are derived by the ground atoms in the extensional database (as opposed to the non-deterministic version of the algorithm that guesses applicable substitutions). This enables us to extend {\tt DeterministicWSQAns} for finding answers to open conjunctive queries, by building resolution proof schemas for all  possible substitutions.

{\tt WeaklyStickyQAns}  runs in polynomial time in the size of the extensional database~\cite{ACL10-2}. It can be proved that {\tt DeterministicWSQAns} also runs in polynomial time.
None of these algorithms are first-order (FO) query rewriting algorithms, which do exist for the Datalog$\pm$ more restrictive syntactic classes, e.g. {\em linear} and {\em sticky} \cite{ACL09-2,ACL10-2}.

The MD ontologies to which the complexity results and algorithms above apply support both upward and downward navigation. However, for simpler MD ontologies that support only upward navigation (which can be syntactically detected from the form of the dimensional rules), we developed a methodology for conjunctive query answering based on FO query rewriting. The rewritten query can be posed directly to the extensional database. Ontologies of this kind are common and natural in real world applications (Example \ref{exp:intr} shows such a case).
Interestingly, these ``upward-navigating" MD ontologies {\em do not} necessarily fall into any of the ``good" cases of Datalog$\pm$ mentioned above.

The algorithms mentioned in this section are rather proofs of concept than algorithms meant to be used with massive data. It is ongoing work the development and implementation of scalable polynomial
time algorithms for answering open conjunctive queries.

\vspace{-2mm}
\section{MD Ontologies and Data Quality} \label{sec:cdqa}

\begin{figure}
\center
\includegraphics[width=6cm]{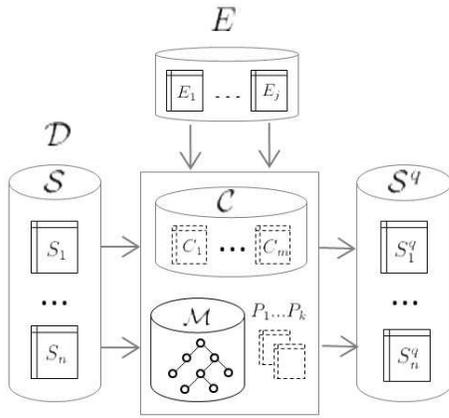}
\caption{An MD context  for data quality assessment}
\label{fig:frmw} \vspace{-4mm}
\end{figure}

\vspace{-1mm}In this section, we show how a Datalog$\pm$ MD ontology can be a part of -and used in- a context for data quality assessment or cleaning. Fig. \ref{fig:frmw} shows such a context and the way it is used. The central idea in \cite{BR10} is that the original instance $\mc{D}$ (on the left-hand-side) is to be assessed or cleaned through the context in the middle. This is done by mapping $\mc{D}$ into the contextual schema/instance $\mc{C}$. The context may have additional data, predicates ($C_i$), data quality predicates ($P_i$) specifying single quality requirements, and access to external data sources ($E_i$) for data assessment or cleaning. The clean version of $\mc{D}$ is on the right-hand-side, with schema $\mc{S}^q$, which is a copy of $\mc{D}$'s  schema \cite{BR10}.

The new element in the context is the MD ontology $\mc{M}$, which interacts with $\mc{C}$, and represents the dimensional elements of the context. The categorical relations in $\mc{M}$ provide dimensional data for the relations in $\mc{C}$ and for quality predicates in $\mc{P}$. $\mc{C}$ also gets extensional data from initial database, $\mc{D}$, and external sources. Here we concentrate on data cleaning, which here amounts to obtaining clean answers to queries, in particular, about clean extensions ($S^q_i$) for the original database relations ($S_i$) (a particular case of {\em clean query answering} \cite{BR10}).

The quality versions $S^q_i$  are specified in terms of the relations in $\mc{C}$ and quality predicates, $P_i$. The data for the latter may be already in the context or come from
$\mc{D}$, the ontology $\mc{M}$, or external sources. The problems become: (a) computing quality versions $S_i^q$ of the original predicates, and (b) computing quality answers to queries $\mc{Q}$ expressed in terms of
those original predicates. The second problem is solved by
rewriting the query as $\mc{Q}^q$, which is expressed (and answered) in terms of predicates $S^q_i$. Answering it is the part of the query answering process that may invoke  dimensional navigation and data generation
as illustrated in previous sections. Problem (a) is a particular case of (b).

\begin{example}(ex.~\ref{exp:ont} cont.) A query $\mc{Q}$ about Tom Waits' temperatures is initially expressed
in terms of the initial predicates \nit{Measurements},  but is rewritten into  a query expressed  and an-

 \newpage \vspace*{-6mm}\noindent swered via its quality extension $\nit{Measurements}^q$ (see \cite{BR10} for more details).\footnote{This idea of cleaning data on-the-fly is reminiscent of {\em consistent query answering} \cite{bertossi11}.}
More specifically, the  query is about \ {\it ``The body temperatures of Tom Waits on September 5 taken around noon by a certified nurse with a thermometer of brand B1"}: \vspace{-2mm}
\begin{align*}
\mc{Q}(t,p,v) \ \leftarrow& \ {\it Measurements(t,p,v)},p={\tt Tom\;\;Waits},\\
& ~~{\tt Sep/5\mbox{-}11\mbox{:}45} \le t \le {\tt Sep/5\mbox{-}12\mbox{:}15}.
\end{align*}
\vspace{-6mm}\phantom{poto}

{\it Measurements}, as initially given, does not contain information about nurses or thermometers. Hence the {\em expected conditions} are not expressed
in the query. According to the general contextual approach in \cite{BR10},  predicate  {\it Measurement} has to be
logically connected to the context, conceiving it as a footprint of a ``broader" contextual relation that is given or built in the context, in this case
one with information about thermometer brands ($b$) and nurses' certification status ($y$):\vspace{-2mm}
\begin{eqnarray*}
{\it Measurement}^{\prime}(t,p,v,y,b)&\leftarrow&{\it Measurement}^c(t,p,v),\\
&&\hspace*{-3.8cm}{\it TakenByNurse}(t,p,n,y), {\it TakenWithTherm}(t,p,b),
\end{eqnarray*}
\phantom{poto}

\vspace{-6mm}\noindent where ${\it Measurement}^c$ is a contextual copy of  {\it Measurement}, i.e. the latter is mapped into the context.\footnote{It does not have to be a replica; it could also be mapped into a contextual relation having additional attributes and data \cite{BR10}.}
If we want quality measurements data, we impose the quality conditions: \vspace{-2mm}
\begin{eqnarray*}
{\it Measurement}^q(t,p,v)&\leftarrow&{\it Measurement}^{\prime}(t,p,v,y,b), \\
&&y={\tt Certified}, \ b={\tt B1},
\end{eqnarray*}
\phantom{poto}

\vspace{-6mm}\noindent with the auxiliary predicates defined by:\vspace{-2mm}
\begin{eqnarray*}
\hspace*{-1mm}{\it TakenByNurse}(t,p,n,y)\hspace{-2mm}&\leftarrow&\hspace{-3mm}{\it WorkingSchedules}(u,d;n,y), \\
&&\hspace{-1.3cm}{\it DayTime}(d,t), {\it PatientUnit}(u,d;p).\\
{\it TakenWithTherm}(t,p,b)&\leftarrow&{\it PatientUnit}(u,d;p),\\
&&\hspace*{-1.8cm}{\it DayTime}(d,t),b={\tt B1},u={\tt Standard}.
\end{eqnarray*}
\phantom{poto}

\vspace{-6mm}\noindent
Here, {\it DayTime} is parent/child relation in {\em Time} dimension), and the last definition
right above is capturing as a rule the guideline from Example~\ref{exp:intr}, at the level of
relation {\it PatientUnit}.

Summarizing, {\it TakenByNurse} and {\it TakenWithTherm} are contextual  predicates (shown in Fig. \ref{fig:frmw} as $P_i$). {\it PatientWard} and {\it WorkingSchedules} are categorical relations.

To obtain quality answers to the original query, we pose to the ontology the new query:\vspace{-2mm}
\begin{eqnarray*}
\mc{Q}^q(t,p,v) \ \leftarrow& \ {\it Measurements(t,p,v)^q}, \ p={\tt Tom\;\;Waits},\\
& {\tt Sep/5\mbox{-}11\mbox{:}45} \le t \le {\tt Sep/5\mbox{-}12\mbox{:}15}.
\end{eqnarray*}
\phantom{poto}

\vspace{-6mm}\noindent
Answering it, which requires evaluating {\it TakenWithTherm}, triggers upward dimensional navigation from {\it Ward} to {\it Unit}, when requesting data for categorical relation {\it PatientUnit}.
More specifically,  dimensional rule (\ref{frm:upward}) is used for data generation, and
each tuple in {\it PatientWard} generates one tuple in {\it PatientUnit}, with its unit obtained by rolling-up .
\boxtheorem\end{example}

\vspace{-3mm}
\section{Conclusions} \label{sec:conc}

\vspace{-1mm}We have described in general terms how to specify in Datalog$\pm$ a multidimensional ontology that extends a multidimensional data model. We have identified some
properties of these ontologies in terms of membership to known classes of Datalog$\pm$, the complexity of conjunctive query answering, and the existence of algorithms
for    the latter task. Finally, we showed how to apply the ontologies to multidimensional and contextual data quality, in particular, for obtaining quality answers to
queries through dimensional navigation. MD contexts are also of interest outside applications to data quality. They can be seen as logical extensions of the MD data model.

\ignore{we have formalized a MD data model first introduced in~\cite{BR10} that extends HM data model~\cite{HG05} with a new class of relations called categorical categorical relations. We have also introduced dimensional rules and constraints on such a data model to address the incompleteness of data in different levels.

We use Datalog$\pm$ ontological language for representing these dimensional rules and constraints and we show that the result ontology has syntactic properties of WS Datalog$\pm$~\cite{ACL12} ontology. This implies that query answering can be done using the query answering algorithm for WS Datalog$\pm$ ontology. Finally, we have applied the MD ontology in a data quality assessment framework in which quality query answering can be done using query answering on the MD ontology.}

\ignore{Ongoing and future research efforts are centered around improving the algorithms for query answering and their implementations. Experimentation with them is also one of our
current concerns.}

\ignore{, we intend to explore more general syntactic forms of dimensional rules while preserving the tractability of query answering in the ontology. In a different direction, we want to improve the performance of query answering in the MD ontology by designing a rewriting method using common query answering approaches such as Magic-set.}

\vspace{1mm}\noindent {\small {\bf Acknowledgments:} Research funded by NSERC Discovery, and the NSERC Strategic Network on Business Intelligence (BIN). L. Bertossi is a Faculty Fellow of IBM CAS. We thank Andrea Cali and Andreas Pieris for useful information and conversations on Datalog$\pm$.}

\bibliographystyle{IEEEtran}

\ignore{
\appendix
\section{Appendix}

\subsection{Algorithms}

\subsubsection{The {\tt DeterministicWSQAns} Algorithm}\label{sec:alg}
The {\tt DeterministicWSQAns} algorithm is a deterministic version of the alternating \texttt{WeaklyStickyQAns} algorithm proposed in \cite{ACL12}. The \texttt{WeaklyStickyQAns} algorithm builds a {\em accepting resolution proof schema} for a query $q$ with respect to an extensional database $D$ and a set of WS TGDs, $\Sigma$. A resolution proof schema is a tree-like structure that is part of the chase for $\Sigma$ and $D$. Such a proof schema shows how $D$ and $\Sigma$ entail $q$.

The {\tt DeterministicWSQAns} algorithm builds an accepting resolution proof schema for $q$, $\Sigma$ and $D$ by a top-down approach, depth first search process. If such a proof exists the algorithm finds the proof in polynomial time (with respect to the size of $D$) and returns true; otherwise it rejects and returns false. The following pseudo-code has the details of the algorithm.

\begin{verbatim}
Algorithm DeterministicWSQAns('D','T','Q')

begin algorithm
comment: decides whether boolean conjunctive query 'Q'
         can be entailed from database 'D' and set of TGDs 'T'

initialize 'K' to an empty stack,
           'A' to the list of atoms in body of 'Q'
           'S' to an empty substitution

push a dummy on 'K'

while 'K' is not empty
    if all atoms in 'A' are resolved
        return true
    initialize 'a' to the first unresolved atom in 'A'
    if there is a resolved atom 'b' var('Q')-isomorphic with 'a'
        if 'a' is generated from expanding 'b'
            rollback to the last decision in 'K'
        else
            update 'S' with substitution that maps 'a' to 'b'
            mark 'a' as resolved
            store information about resolving 'a' on 'K'
    if there exists a not-applied substitution 'U' that maps 'a' to 'D'
        update 'S' with 'U'
        mark 'a' as resolved
        store information about resolving 'a' on 'K'
    else if there exists a not-applied rule 't' in 'T' that can resolve 'a'
        if 'a' and 't' make a wrong critical node
            rollback to the last decision in 'K'
                that caused the corresponding valid critical node
        else
            update 'S' with a substitution that maps head of 't' to 'a'
            expand 'A' replacing 'a' with body of 't'
            mark 'a' as resolved
            store information about resolving 'a' on 'K'
    else
        rollback to the last decision in 'K'

return false
end algorithm
\end{verbatim}

For a query $q$, a set of dimensional rules $\Sigma$, and a database instance $D$, \texttt{DeterministicWSQAns} resolve the atoms in $q$ from left to right. An atom is resolved by (1) finding a substitution that maps the atom to a ground atom (substitution is a homomorphism that maps variables to constants) in $D$, (2) applying a rule (that results to a subtree with leaf nodes in $D$). In each step (implemented in the code with an iteration of the main while loop), an atom from the list of atoms is resolved. The current substitution and the pointer to current atom are stored on a stack and then updated according to the decision that is made in the current step. If an atom can not be resolved, the algorithm pops the last decision (values for the pointer and the substitution) from the stack and continues with the last values stored on the stack (roll-back phase).

Note that if the current atom is {\it var(q)-isomorphic} ({\it vat(q)} is the set of variables in the given query $q$) with a previously resolved atom then the current atom will be marked as resolved if the atoms appear in two different subtrees in the resolution proof schema (that is built so far). But if the current atom and its {\it var(q)-isomorphic} atom are in the same subtree, the algorithm rolls-back.

\subsection{Proofs}\label{sec:proofs}
\defproof{Theorem~\ref{th:t1}}{ A set of TGDs, $\Sigma$ over a schema $\mc{R}$ is weakly sticky if and only if for each $\sigma \in \Sigma$, and for each variable $V$ that occurs more than once in ${\it body}(\sigma)$, the following holds: $V$ is {\it non-marked} (for definition of marked variables see~\cite{ACL12}), or at least one occurrence of $V$ in ${\it body}(\sigma)$ appears at some positions of $\Pi_F(\mc{R},\Sigma)$ ($\Pi_F(\mc{R},\Sigma)$ is the set of positions in the dependency graph of $\Sigma$ that does not appear in a cycle that goes through existential (special) edges. Intuitively, $\Pi_F(\mc{R},\Sigma)$ contains positions that limited number of constants or nulls can appear in them during the chase procedure~\cite{FG03}).

Consider the dependency graph for a the set of dimensional rules ($\Sigma$). Based on the definition of dimensional rules specified by Rule~\ref{frm:gf4}, categorical attributes refer to positions in $\Pi_F$ (no existentially quantified variable occurs at these positions or the positions in the dependency graph that have a path to these positions). Since the shared variables in the body of these dimensional rules only occur in positions of categorical attributes, we can conclude that all the shared variables are in $\Pi_F$. So, a set of dimensional rules (of the form Rule~\ref{frm:gf4}) has the syntactic property of a WS set of TGDs.}  }

\end{document}